\pdfoutput=1
\documentclass[aps,prd,preprint,superscriptaddress,tightenlines,nofootinbib]{revtex4}

\long\def\symbolfootnote[#1]#2{\begingroup%
\def\thefootnote{\fnsymbol{footnote}}\footnote[#1]{#2}\endgroup}

\usepackage{amsfonts,amsmath,amsthm,amssymb}
\usepackage{mathrsfs}
\usepackage{bm}
\usepackage{color}
\usepackage{epsfig}

\newcommand{\PRE}[1]{{#1}}   

\newcommand{\beq}{\begin{equation}}
\newcommand{\eeq}{\end{equation}}
\newcommand{\bea}{\begin{flushleft} \begin{eqnarray}}
\newcommand{\eea}{\end{eqnarray}\end{flushleft}}

\newcommand{\postscript}[2]{\setlength{\epsfxsize}{#2\hsize}
   \centerline{\epsfbox{#1}}}
\newcommand{\comment}[1]{}

\newcommand{\ci}[1]{}

\newcommand{\lb}{\left(}
\newcommand{\rb}{\right)}

\newcommand{\ba}{\begin{eqnarray}}
\newcommand{\ea}{\end{eqnarray}}
\newcommand{\be}{\begin{equation}}
\newcommand{\ee}{\end{equation}}
\newcommand{\bay}[1]{\left(\begin{array}{#1}}
\newcommand{\eay}{\end{array}\right)}

\newcommand{\zt}[1]{\rm{#1}}

\def\xg{{\gamma}}
\def\xG{{\Gamma}}

\def\xvp{{\varphi}}
\def\xs{{\sigma}}

\usepackage[usenames,dvipsnames]{xcolor}
\definecolor{orange}{cmyk}{0,0.5,1,0}
\definecolor{rossoCP3}{cmyk}{0,.88,.77,.40}
\definecolor{graa}{rgb}{0.8,0.8,0.8}
\definecolor{blaa}{rgb}{0.2,0.2,0.6}

\begin{document}

\preprint{
\hfil
\begin{minipage}[t]{3in}
\begin{flushright}
\vspace*{.4in}
LMU-ASC 13/16\\
\end{flushright}
\end{minipage}
}

\title{\PRE{\vspace*{0.9in}} \color{rossoCP3}{Update on 750~GeV diphotons from
    closed string states 
}}

\author{{\bf Luis A. Anchordoqui}}

\affiliation{Department of Physics and Astronomy,\\  Lehman College, City University of
  New York, NY 10468, USA
\PRE{\vspace*{.05in}}
}

\affiliation{Department of Physics,\\
 Graduate Center, City University
  of New York,  NY 10016, USA
\PRE{\vspace*{.05in}}
}

\affiliation{Department of Astrophysics,\\
 American Museum of Natural History, NY
 10024, USA
\PRE{\vspace*{.05in}}
}

\author{{\bf Ignatios Antoniadis}}
\affiliation{LPTHE, UMR CNRS 7589\\
Sorbonne Universit\'es, UPMC Paris 6, 75005 Paris, France
\PRE{\vspace*{.05in}}}

\affiliation{Albert Einstein Center, Institute for Theoretical Physics\\
Bern University, Sidlerstrasse 5, CH-3012 Bern, Switzerland
\PRE{\vspace*{.05in}}}

\author{{\bf Haim \nolinebreak Goldberg}}
\affiliation{Department of Physics,\\
Northeastern University, Boston, MA 02115, USA
\PRE{\vspace*{.05in}}
}

\author{{\bf Xing Huang}}
\affiliation{Department of Physics, \\
National Taiwan Normal University, Taipei, 116, Taiwan
\PRE{\vspace*{.05in}}
}

\author{{\bf Dieter L\"ust}}

\affiliation{Max--Planck--Institut f\"ur Physik, \\ 
 Werner--Heisenberg--Institut,
80805 M\"unchen, Germany
\PRE{\vspace*{.05in}}
}

\affiliation{Arnold Sommerfeld Center for Theoretical Physics 
Ludwig-Maximilians-Universit\"at M\"unchen,
80333 M\"unchen, Germany
\PRE{\vspace{.05in}}
}

\author{{\bf Tomasz R. Taylor}}

\affiliation{Department of Physics,\\
 Northeastern University, Boston, MA 02115, USA 
 \PRE{\vspace*{.05in}}
}

\begin{abstract}
  \noindent Motivated by the recent update on LHC searches for narrow
  and broad resonances decaying into diphotons we reconsider the
  possibility that the observed peak in the invariant mass spectrum at
  $M_{\gamma \gamma} = 750~{\rm GeV}$ originates from a closed string
  (possibly axionic) excitation $\varphi$ (associated with low mass
  scale string theory) that has a coupling with gauge kinetic
  terms. We reevaluate the production of $\varphi$ by photon fusion to
  accommodate recent developments on additional contributions to
  relativistic light-light scattering. We also study the production of
  $\varphi$ via gluon fusion. We show that for both a narrow and a
  broad resonance these two initial topologies can accommodate the
  excess of events, spanning a wide range of string mass scales $7
  \alt M_s/{\rm TeV} \alt 30$ that are consistent with the
  experimental lower bound: $M_s > 7~{\rm TeV}$, at 95\% CL. We
  demonstrate that for the two production processes the LHC13 data is
  compatible with the lack of a diphoton excess in LHC8 data within
  $\sim 1\sigma$.  We also show that if the resonance production is
  dominated by gluon fusion the null results on dijet searches at LHC8
  further constrain the coupling strengths of $\varphi$, but without
  altering the range of possible string mass scales.

\end{abstract}

\maketitle

Recently, the ATLAS~\cite{ATLAS} and CMS~\cite{CMS:2015dxe}
collaborations reported excesses of events over expectations from
standard model (SM) processes in the diphoton mass distribution around
750~GeV, using (respectively) $3.2~{\rm fb}^{-1}$ and $2.6~{\rm
  fb}^{-1}$ of data recorded at a center-of-mass energy $\sqrt{s} =
13~{\rm TeV}$.  This could be interpreted as decays of a new massive
particle $\varphi$. For a narrow width approximation hypothesis, the
ATLAS Collaboration gives a local significance of $3.6\sigma$ and a
global significance of $2.0\sigma$ when accounting for the
look-elsewhere-effect in the mass range $M_\varphi/{\rm GeV} \in [200
- 2000]$. Signal-plus-background fits were also implemented assuming a
large decay width for the signal component. The most significant
deviation from the background-only hypothesis is reported for
$M_\varphi \sim 750~{\rm GeV}$ and a total width \mbox{$\Gamma_{\rm
    total} \approx 45~{\rm GeV}$.} The local and global significances
evaluated for the large width fit are about 0.3 higher than that for
the fit using the narrow width approximation, corresponding to
$3.9\sigma$ and $2.3\sigma$, respectively. The CMS data yields a local
significance of $2.6\sigma$ and a global significance smaller than
$1.2\sigma$. Fitting the LHC13 data with a resonance yields a cross
section times branching ratio of 
\begin{equation}
\sigma_{\rm LHC13} (pp \to \varphi +
{\rm 
anything}) \times {\cal B} (\varphi \to \gamma \gamma) \approx
\left\{\begin{array}{cl} 
(10 \pm 3)~{\rm fb} &~~~ {\rm ATLAS} \\
\phantom{0}(6 \pm 3)~{\rm fb} &~~~ {\rm CMS} \end{array}
\right. \,,
\end{equation}
at $1\sigma$~\cite{Franceschini:2015kwy}. On the other hand, no diphoton resonances
were seen in the data at $\sqrt{s} = 8~{\rm TeV}$, although both
ATLAS~\cite{Aad:2015mna} and CMS~\cite{Khachatryan:2015qba} data show
a mild upward fluctuation at invariant mass of 750~GeV.  The lack of
an excess at $\sqrt{s} = 8~{\rm TeV}$ allows a quite precise limit to
be placed on the corresponding cross section at $\sqrt{s} = 13~{\rm
  TeV}$. The most stringent limit comes from the CMS search
$\sigma_{\rm LHC8} (pp \to \varphi + {\rm anything}) \times {\cal B}
(\varphi \to \gamma \gamma) < 2.00~{\rm fb}$ at 95\%
CL~\cite{Khachatryan:2015qba}.  This implies that if the diphoton
cross section grows by less than about a factor of 3 or 3.5 the LHC8 data are
incompatible with the LHC13 data at 95\% CL.

More recently, we proposed a model~\cite{Anchordoqui:2015jxc} to
explain the data in which the resonance production mechanism is
calculable in string based dynamics, with large extra
dimensions~\cite{Antoniadis:1998ig}.  In our proposal the observed
diphoton excess originates from a closed string excitation $\xvp$
living on the compact space of generic intersecting D-brane models
that realize the SM chiral matter contents and gauge
symmetry~\cite{Blumenhagen:2005mu,Blumenhagen:2006ci}.  There are two
properties of the scalar $\varphi$ that are necessary for explaining
the 750~GeV signal. It should be a special closed string state with
dilaton-like or axion-like coupling to $F^2$ (respectively to $F
\tilde F$) of the electromagnetic field, but {\em may be decoupled}
from $G^2$ of color $SU(3)$.  The couplings of closed string states to
gauge fields do indeed distinguish between different D-brane stacks,
depending on the localization properties of D-branes with respect to
$\varphi$ in the compact dimensions.  More specifically, it is quite
natural to assume that $\varphi$ is a closed string mode that is
associated to the wrapped cycles of the (lepton) $U(1)_L$ and (right
isospin) $U(1)_{I_R}$ stack of D-branes, however is not or only weakly
attached to the wrapped cycle of (left) $Sp(1)_L$ or the color $SU(3)$
stack of D-branes. In this way, we may avoid unwanted dijet
signals. Actually, within a selection of string based explanations of
the
resonance~\cite{Heckman:2015kqk,Cvetic:2015vit,Ibanez:2015uok,Palti:2016kew,Karozas:2016hcp,Faraggi:2016xnm,Anastasopoulos:2016cmg,Cvetic:2016omj,Li:2016tqf,Leonatris}
our proposal is uniquely exemplified by the {\it possible
  suppression} of dijet topologies in the final
state.\footnote{A related stringy explanation in which
  $\varphi$ can be produced through photon fusion has been put forward
  in~\cite{Abel:2016pyc}.  Alternative axion and dilaton models have
  been discussed in~\cite{Higaki:2015jag,Megias:2015ory,Ben-Dayan:2016gxw,Barrie:2016ntq,Aparicio:2016iwr}.} By choice,
as we already advertised in~\cite{Anchordoqui:2015jxc}, we may also allow a
coupling $\varphi$ to $G^2$. This is possible by modifying the
localization properties of D-branes with respect to $\varphi$ in the
internal space. 

Very recently, the ATLAS and CMS Collaborations updated their diphoton
resonance searches~\cite{Delmastro,Musella,CMS:2016owr}. The ATLAS Collaboration
reported two separate analyses performed with $3.2~{\rm fb}^{-1}$ of data at
13 TeV, targeting spin-0 and spin-2 resonances. For spin-0, the
largest deviation from the background-only hypothesis is reported for
$M_\varphi \sim 750~{\rm GeV}$ and $\Gamma_{\rm total} \approx 45~{\rm
  GeV}$. While the local significance somewhat increases to
$3.9\sigma$ the global significance remains at the $2 \sigma$
level. For the spin-2 resonance, both the local and global
significances are somewhat smaller: $3.6\sigma$ and $1.8\sigma$,
respectively. The new CMS analysis is based on 3.3~${\rm fb}^{-1}$
collected at $\sqrt{s} = 13~{\rm TeV}$. The additional data was
recorded in 2015 while the CMS magnet was not operated. The largest
excess is observed for $M_\varphi = 760~{\rm GeV}$ and $\Gamma_{\rm
  total} \approx 11~{\rm GeV}$ and has a local significance of
$2.8\sigma$ for spin-0 and $2.9\sigma$ spin-2 hypothesis. After taking
into account the effect of searching for several signal hypotheses,
the significance of the excess is reduced to $< 1 \sigma$. The CMS
Collaboration also reported a combined search on data collected at
$\sqrt{s} = 13~{\rm TeV}$ and $\sqrt{s} = 8~{\rm TeV}$. For the
combined analysis, the largest excess is observed at $M_\varphi =
750~{\rm GeV}$ and $\Gamma_{\rm total} = 0.1~{\rm GeV}$. The local
significance is $\approx 3.4\sigma$ and the global significance $1.6\sigma$.

In this short note we extend our previous discussion in three
directions.  The first is a calculation to accommodate recent
developments on additional contributions to relativistic light-light
scattering~\cite{Csaki:2016raa,Harland-Lang:2016qjy,Martin:2016byf}.
The second is the explicit calculation for production via gluon fusion
disclosed in~\cite{Anchordoqui:2015jxc}. The third is a scan of the
parameter space to entertain the possibility of a narrow width favored
by the recent CMS analysis that combines data from LHC8 and LHC13.
Before proceeding we note that the ATLAS excess is quite broad and
probably with a large uncertainty. The CMS excess, however, is smaller and has
no clear preference for a large width. This seems to indicate that the
ATLAS excess could be a real signal combined with a large fluctuation,
making the excess appear larger and wider than the underlying physical
signal. Throughout we  assume the resonance
needs to have a signal~\cite{Kats:2016kuz} 
\begin{equation}
\sigma_{\rm LHC13} (pp \to \varphi + {\rm
  anything}) \times {\cal B} (\varphi \to \gamma \gamma) \approx 3 -
6~{\rm fb}\, .
\label{mattS}
\end{equation}

\begin{figure}[tpb]
\postscript{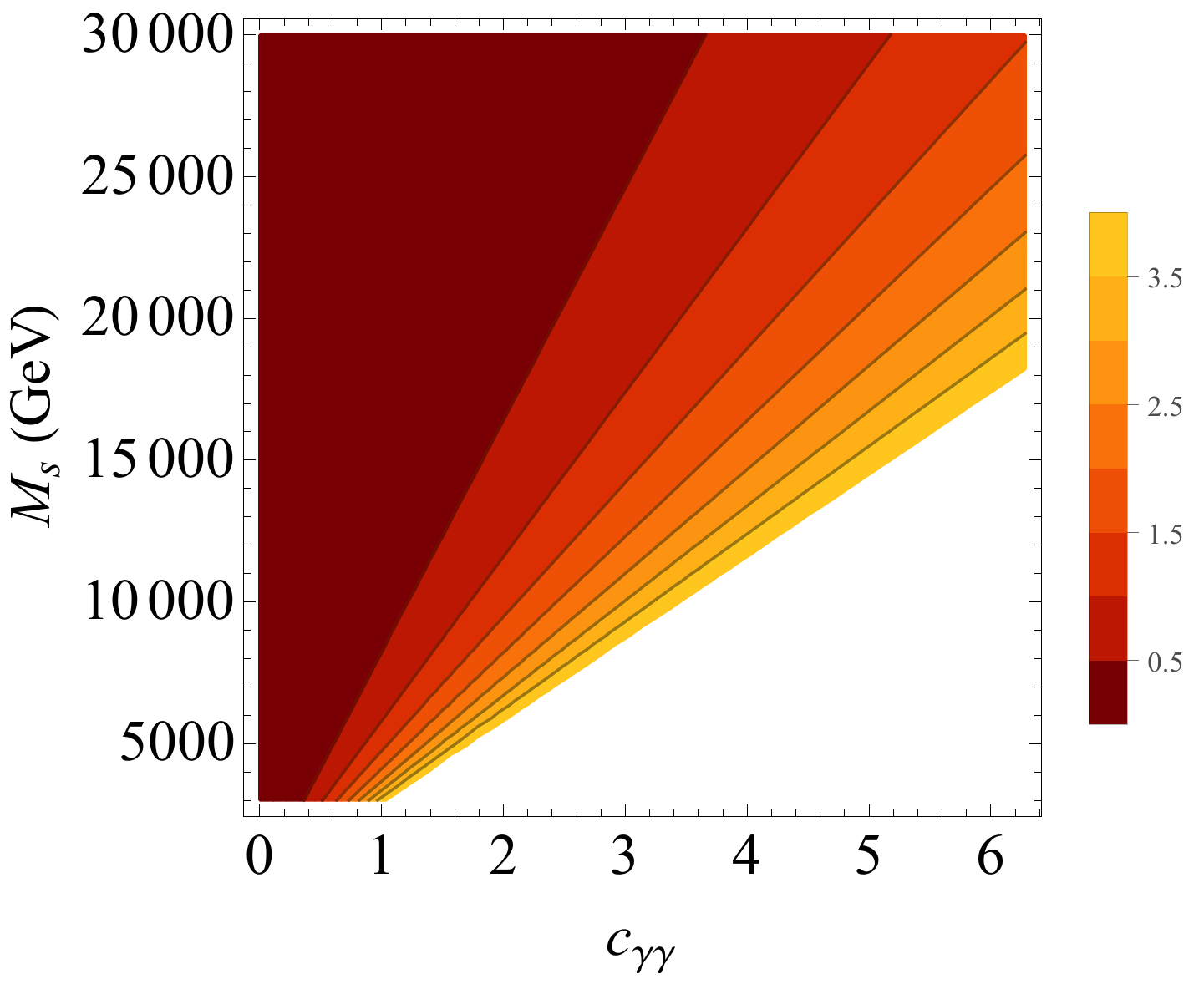}{0.5}
\vspace{-0.3cm}
\caption{Contours of constant partial width $\xG_{\xg\xg}$. The color
  encoded scales are in GeV.}
\label{fig1}
\end{figure}

Since the string mass scale
is now known to be larger than $M_s \approx 7~{\rm
  TeV}$~\cite{Khachatryan:2015dcf}, the mass $M_{\varphi}\approx
750$~GeV must be suppressed with respect to the string scale by some
anomalous loop corrections. Because $\varphi$ is a twisted closed
string localized at an orbifold singularity, its coupling to $\gamma
\gamma$ should be suppressed by $M_s^{-1}$, provided the bulk is
large~\cite{Antoniadis:2002cs}. With this in mind, we parametrize the
coupling of $\xvp$ to the photon by the following vertex \be \frac
{c_{\xg\xg}}{M_s} \, \xvp \, F^2\, . \ee  To remain in the
perturbative range, we also require $c_{\xg\xg} \alt 2 \pi$. The partial decay width of $\xvp$ to diphotons then follows as
\begin{equation}
\Gamma_{\gamma\gamma} = \frac{c_{\gamma\gamma}^2}{4\pi}
\frac{M_\xvp^3}{M_s^2}\, . \label{eq:photonwidth}
\end{equation}
In Fig.~\ref{fig1} we exhibit a scan of the parameter space
 $(c_{\xg\xg}, M_s)$ for constant values of 
 $\Gamma_{\gamma \gamma}$ as obtained from (\ref{eq:photonwidth}).

 Let us first assume the diphoton signal is produced via photon-photon
 fusion with $\xvp$ as the resonance
 state. Following~\cite{Harland-Lang:2016qjy}, herein we include the
 elastic-elastic processes (already considered
 in~\cite{Anchordoqui:2015jxc}) as well as elastic-inelastic and
 inelastic-inelastic contributions. The elastic production is
 suppressed with respect to inelastic by about an order of
 magnitude. However, elastic photoproduction events result in forward
 and backward protons which can be detected by forward detectors
 installed by ATLAS~\cite{ATLAS-let} and CMS~\cite{Albrow}. Therefore,
 the detection of two unbroken protons in the final state, with
 $M_{pp}$ paired to $M_{\gamma \gamma}$, may be a promising way to
 reduce the background in the
 future~\cite{Csaki:2016raa,Martin:2016byf}.

\begin{figure}
 \postscript{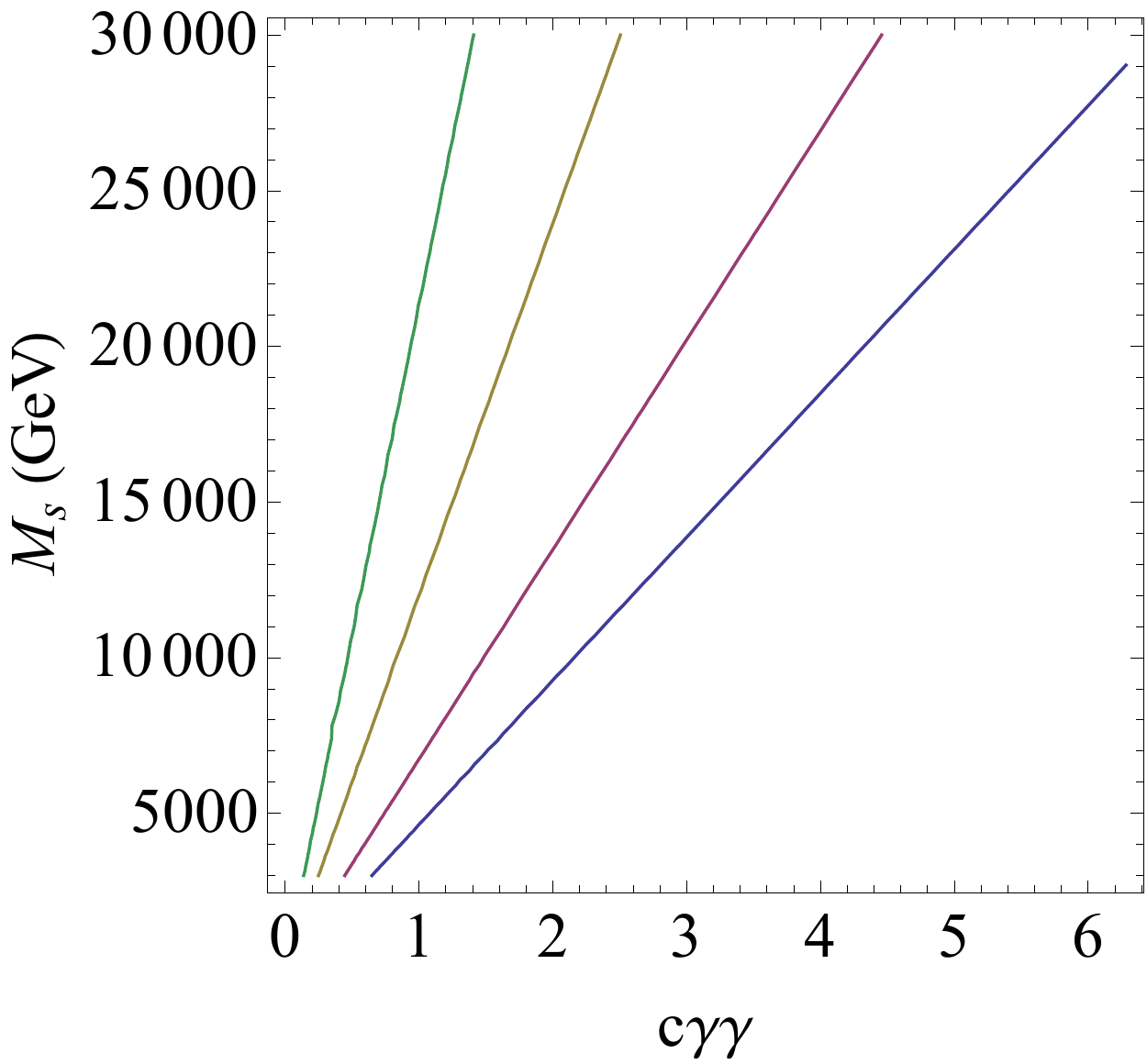}{0.5}
 \caption{Best fit contours of diphoton cross section $\sigma_{\rm
     LHC13} \sim 5~{\rm fb}$ produced via photon fusion $\gamma \gamma
   \to \xvp \to\xg\xg$. The four lines (blue, red,  yellow, green) are
   for $\Gamma_{\zt{total}}= 45, 10, 1.0, 0.1$~GeV, respectively.}
\label{fig2}
\end{figure}

The total photo-production cross section at LHC13
is~\cite{Harland-Lang:2016qjy} \be \sigma_{\rm LHC13} (\gamma \gamma
\to \varphi \to \gamma \gamma) =
4.1~\text{pb}~\left(\frac{\Gamma_{\zt{total}}}{45~\text{GeV}} \right)
{\cal B}^2(\xvp \rightarrow \gamma\gamma)\, ,
\label{shl-1}
\ee where \be {\cal
  B}(\xvp \rightarrow \gamma\gamma) = \frac{2.3 \times 10^6 c_{\gamma
    \gamma}^2}{\pi }\, \left(\frac{M_s}{{\rm GeV}}
\right)^{-2} \, \left(\frac{\Gamma_{\rm total}}{45~{\rm GeV}}
\right)^{-1} . 
\label{shl-2}
\ee Substituting (\ref{shl-2}) into (\ref{shl-1}), and demanding
(\ref{shl-1})  reproduces the diphoton signal (\ref{mattS}) we obtain an
equation connecting $c_{\xg \xg}$ with $M_s$ for given $\Gamma_{\rm
  total}$. In Fig.~\ref{fig2} we show the best fit contours for
$\sigma_{\rm LHC13} \sim 5~{\rm fb}$ and total widths $\Gamma_{\rm
  total} = 45, 10, 1, 0.1~{\rm GeV}$.  We assume that for a broad
resonance the missing fraction of the decay width arises from the
coupling of $\varphi$ to some fermion bulk fields. These hidden
fermions could make a contribution to the dark matter content of the
universe~\cite{Dienes:2011ja,Dienes:2011sa}.  We conclude that for
both the narrow and the broad resonance hypotheses there is an allowed
region of the parameter space which is consistent with the
experimental lower bound of $M_s \simeq 7~{\rm
  TeV}$~\cite{Khachatryan:2015dcf} and reproduces the LHC13 signal.
For the broad resonance hypothesis, $7 \alt M_s/{\rm TeV} \alt 30$.

The total photo-production cross section at LHC8 is~\cite{Harland-Lang:2016qjy}
\be \sigma_{\rm LHC8} (\gamma \gamma \to \varphi \to \gamma \gamma) =
1.4~\text{pb}~\left(\frac{\Gamma_{\zt{total}}}{45~{\rm GeV}} \right)
{\cal B}^2(\xvp \rightarrow \gamma\gamma)\, , \ee
showing consistency with the 95\% CL upper limit~\cite{Khachatryan:2015qba}. Actually, the ratio of
the LHC13/LHC8 partonic luminosity is largely dominated by
systematic uncertainties driven by the parton distribution functions. 
The luminosity ratio is~\cite{Harland-Lang:2016qjy}
\begin{equation}
  \frac{{\cal L}_{\gamma \gamma} (\sqrt{s} = 13~{\rm TeV})}{{\cal L}_{\gamma \gamma} (\sqrt{s} =
    8~{\rm TeV})} = 3^{+0.1}_{-0.2}, \  2.65\pm 0.15, \ 2.1 \pm0.4,
\end{equation}
for CT14QED~\cite{Schmidt:2015zda}, MRST2004~\cite{Martin:2004dh}, and
NNPDF2.3~\cite{Ball:2013hta}; respectively. We note that the predictions
of NNPDF2.3 are only marginally compatible with LHC8
data~\cite{Khachatryan:2015qba}.  

\begin{figure}[tpb]
\postscript{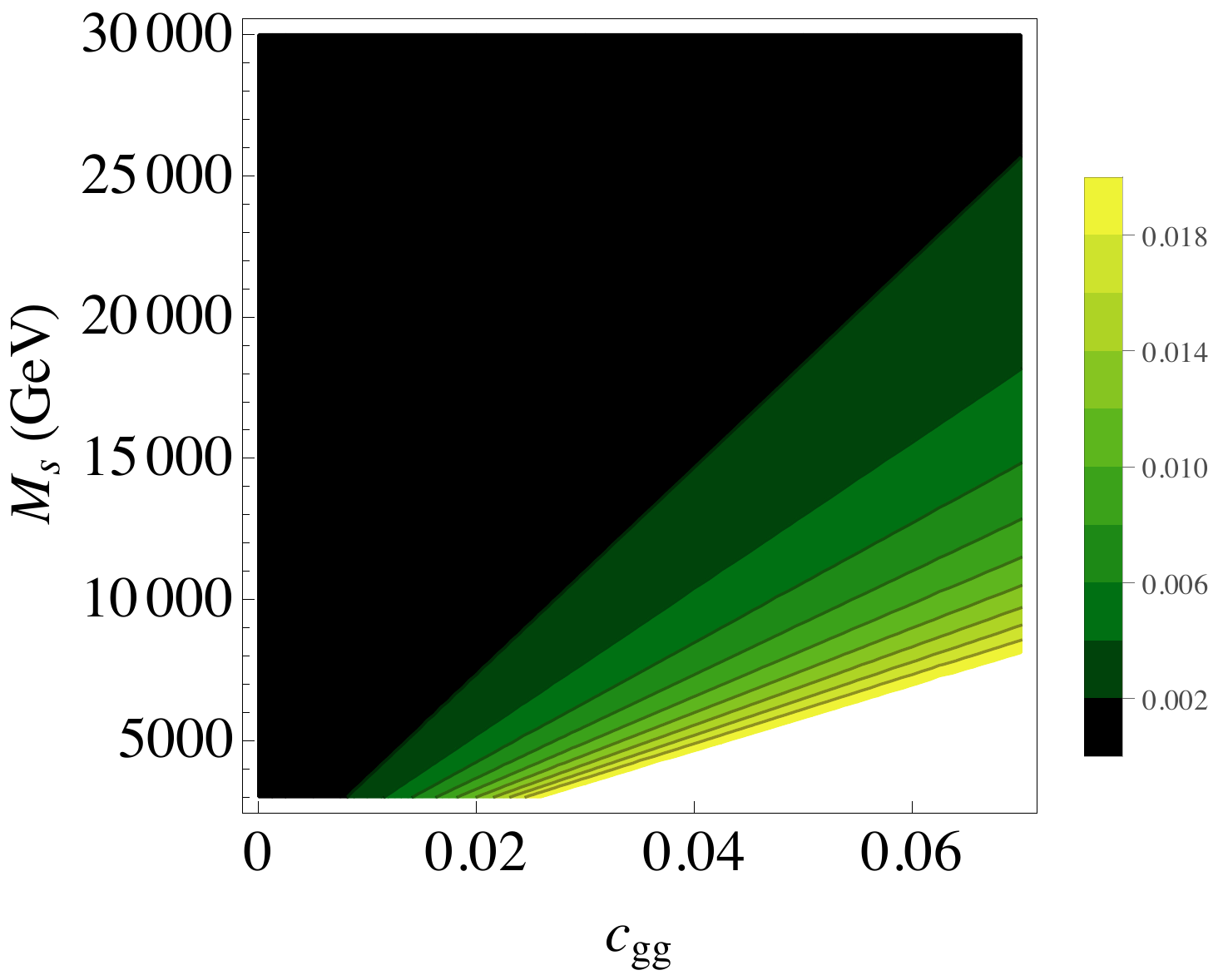}{0.5}
\caption{Contours of constant partial width $\xG_{gg}$. The color
  encoded scales are in GeV.}
\label{fig3}
\end{figure}

\begin{figure}[tpb]
\begin{minipage}[t]{0.7\textwidth}
\postscript{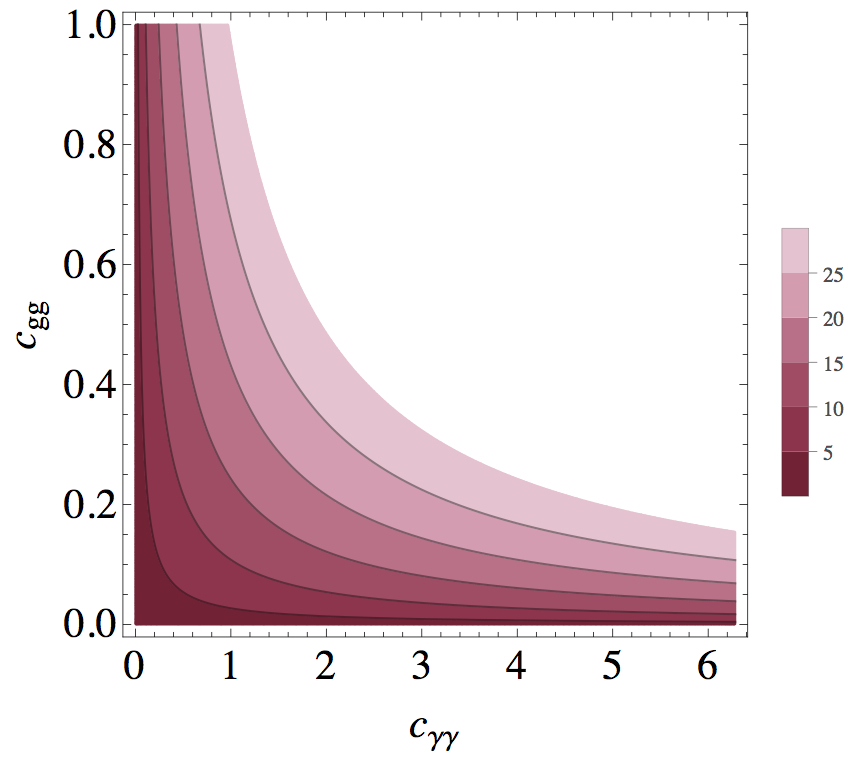}{0.7}
\end{minipage}
\begin{minipage}[t]{0.7\textwidth}
\postscript{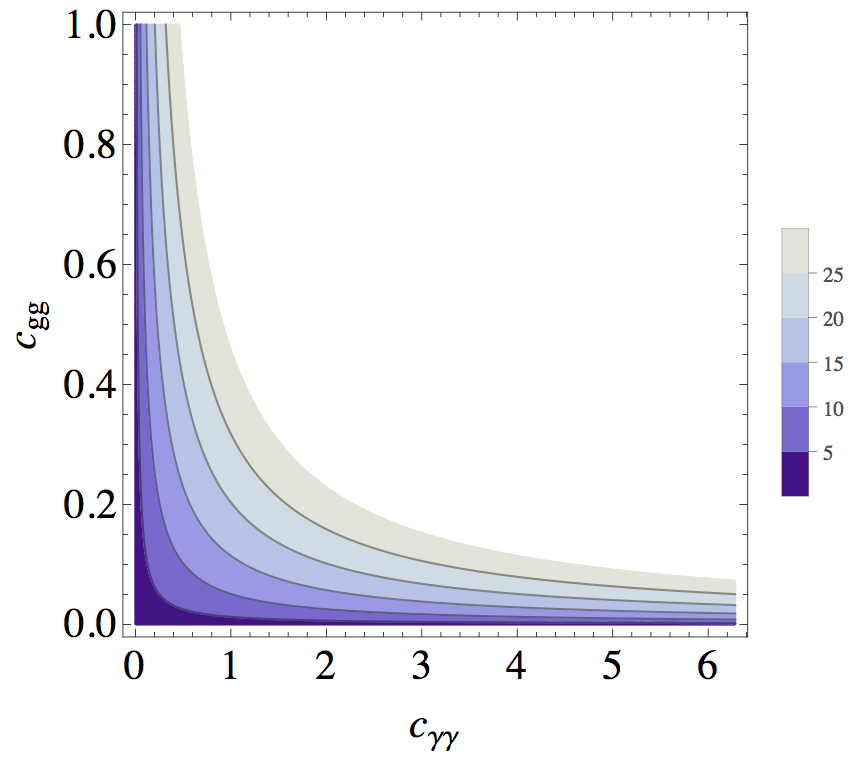}{0.7}
\end{minipage}
\begin{minipage}[t]{0.7\textwidth}
\postscript{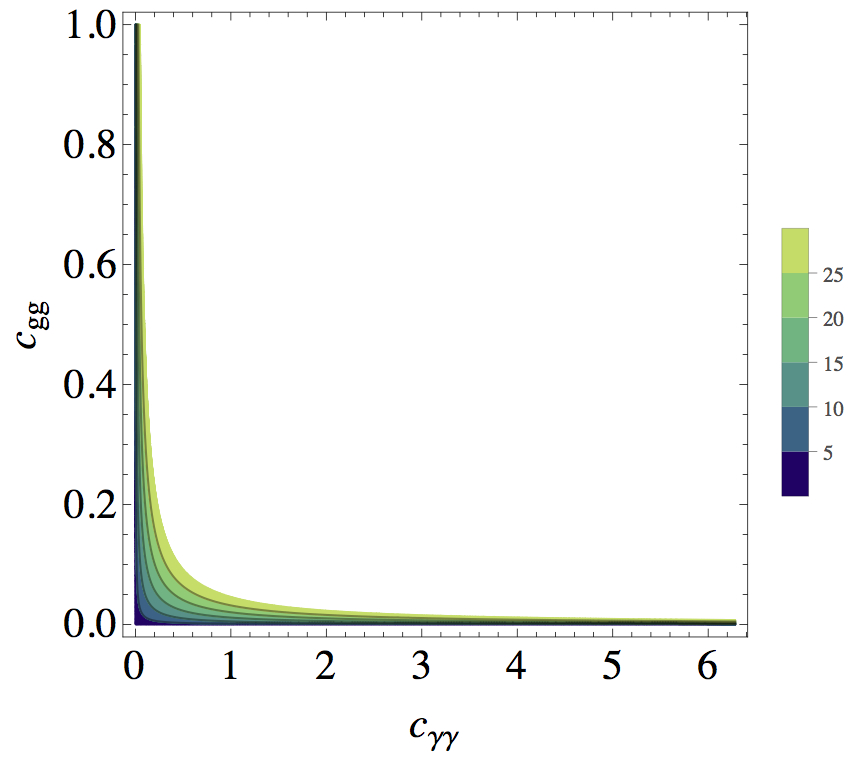}{0.7}
\end{minipage}
\vspace{-0.3cm}
\caption{Contours of constant string scale $M_s$ for best fit of diphoton
  cross section ($\xs_{\rm LHC13} = 5~{\rm fb}$) produced via gluon
  fusion ($M_{\xvp} \simeq 750~{\rm GeV}$, $\sqrt{s} = 13~{\rm
    TeV}$). The color encoded scales are in TeV. The different panels
  correspond to $\xG_{\zt{total}} = 45, 10, 0.1$~GeV, downwards.}
\label{fig4}
\end{figure}

\begin{figure}[tpb]
\begin{minipage}[t]{0.49\textwidth}
\postscript{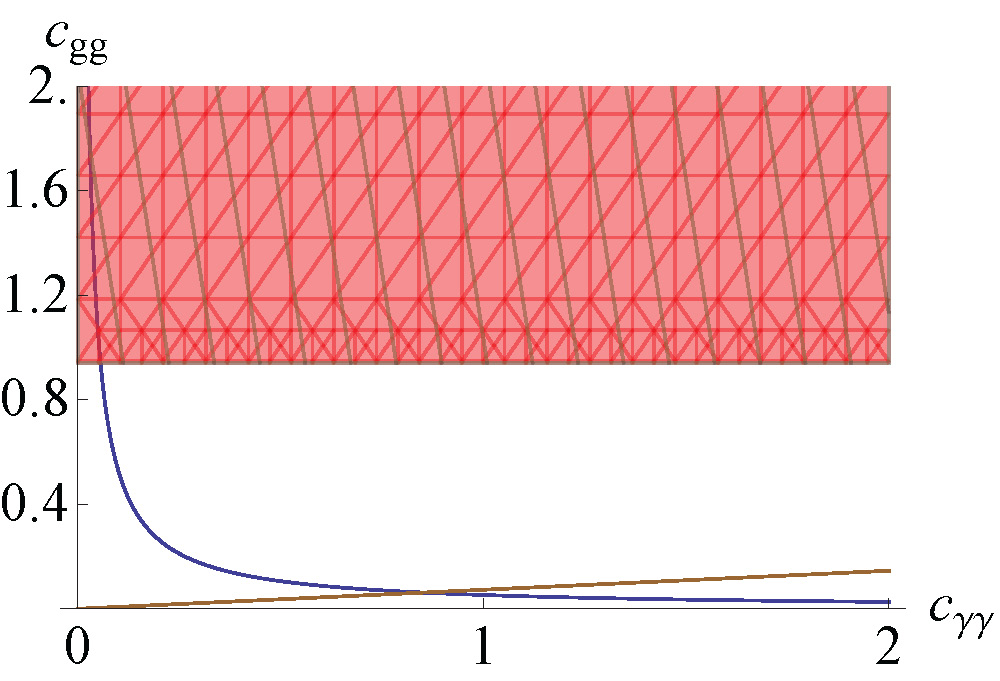}{0.9}
\end{minipage}
\begin{minipage}[t]{0.49\textwidth}
\postscript{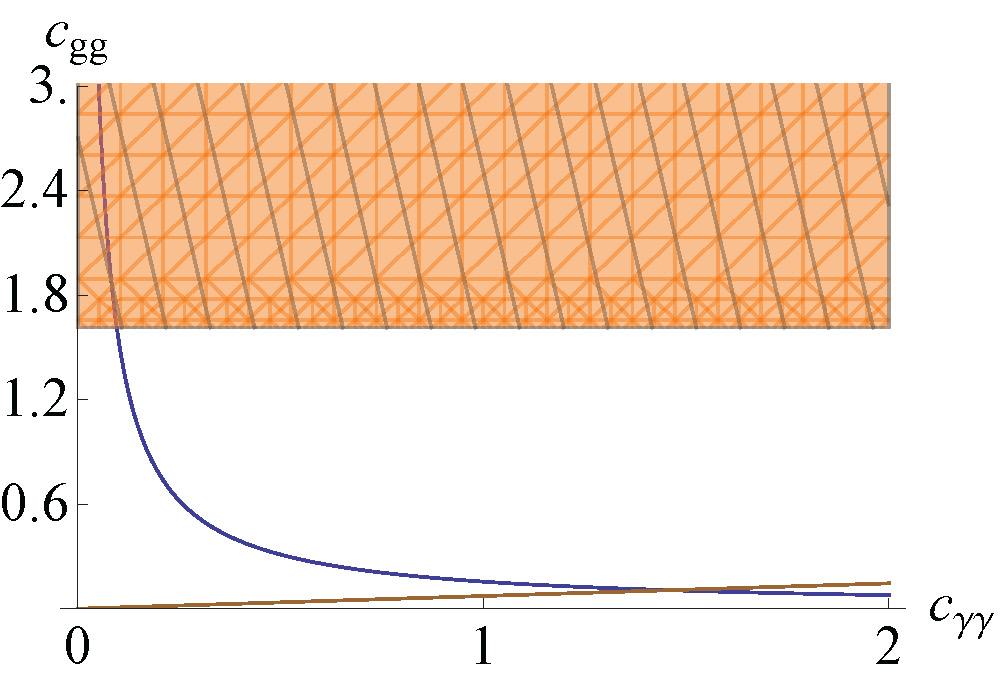}{0.9}
\end{minipage}
\begin{minipage}[t]{0.48\textwidth}
\postscript{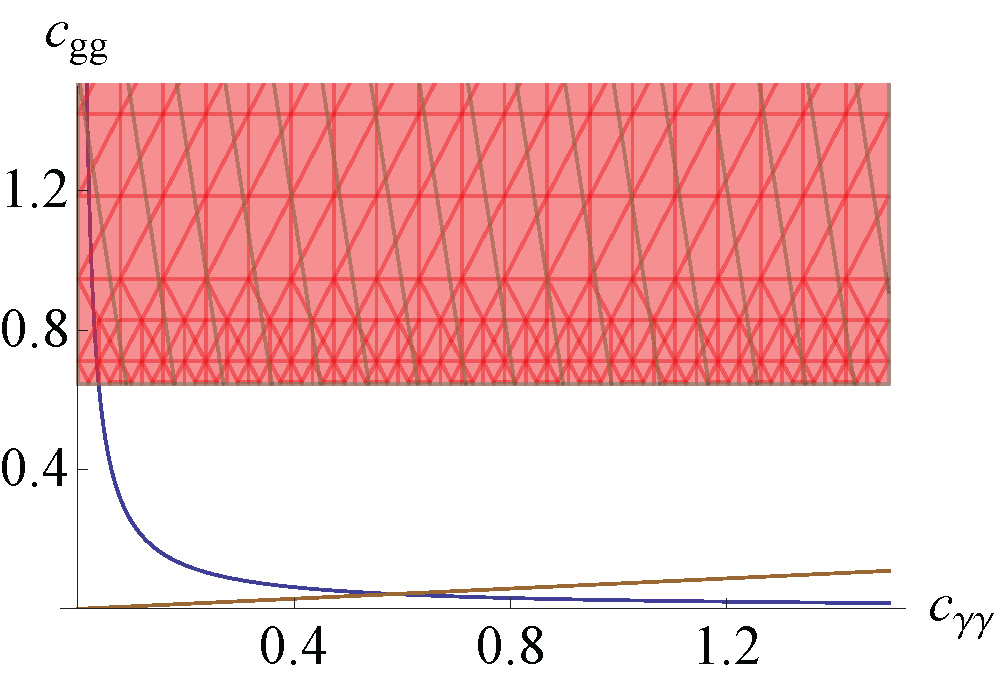}{0.9}
\end{minipage}
\begin{minipage}[t]{0.48\textwidth}
\postscript{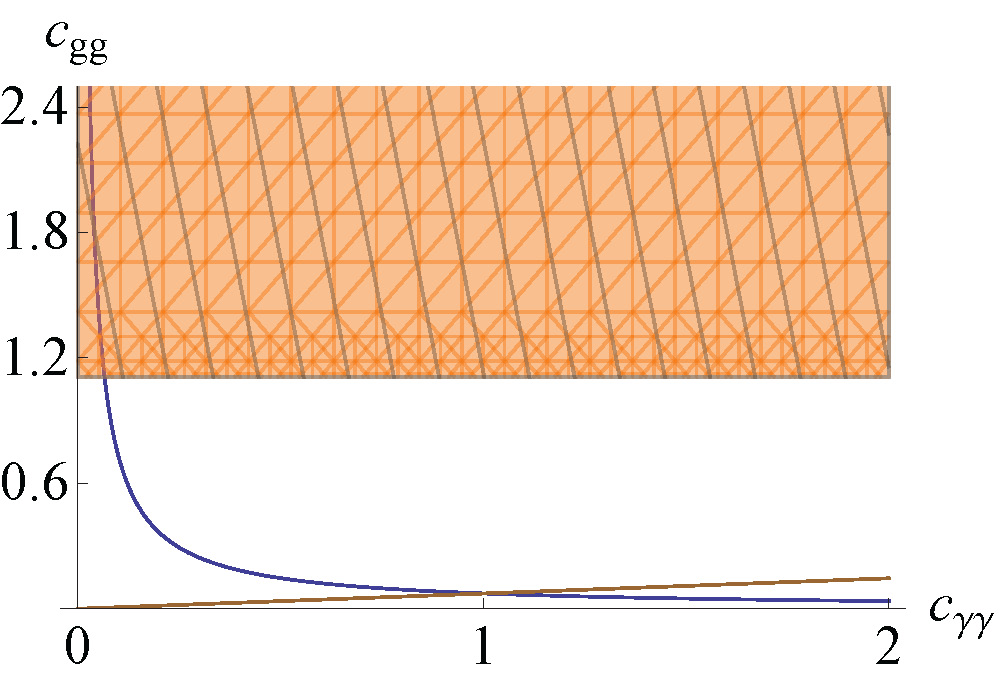}{0.9}
\end{minipage}
\begin{minipage}[t]{0.48\textwidth}
\postscript{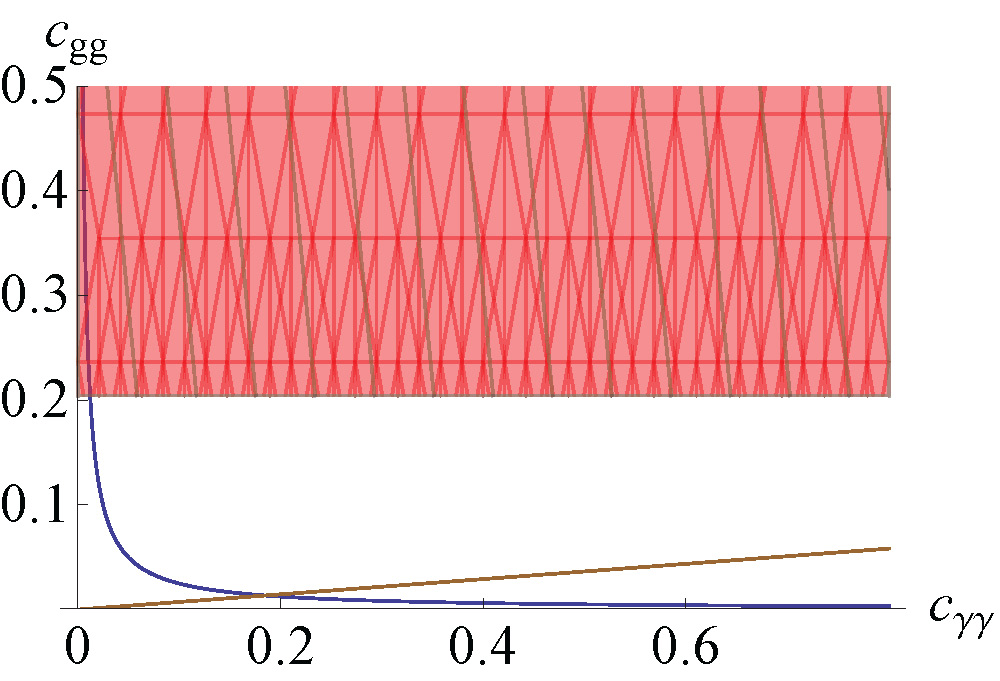}{0.9}
\end{minipage}
\begin{minipage}[t]{0.48\textwidth}
\postscript{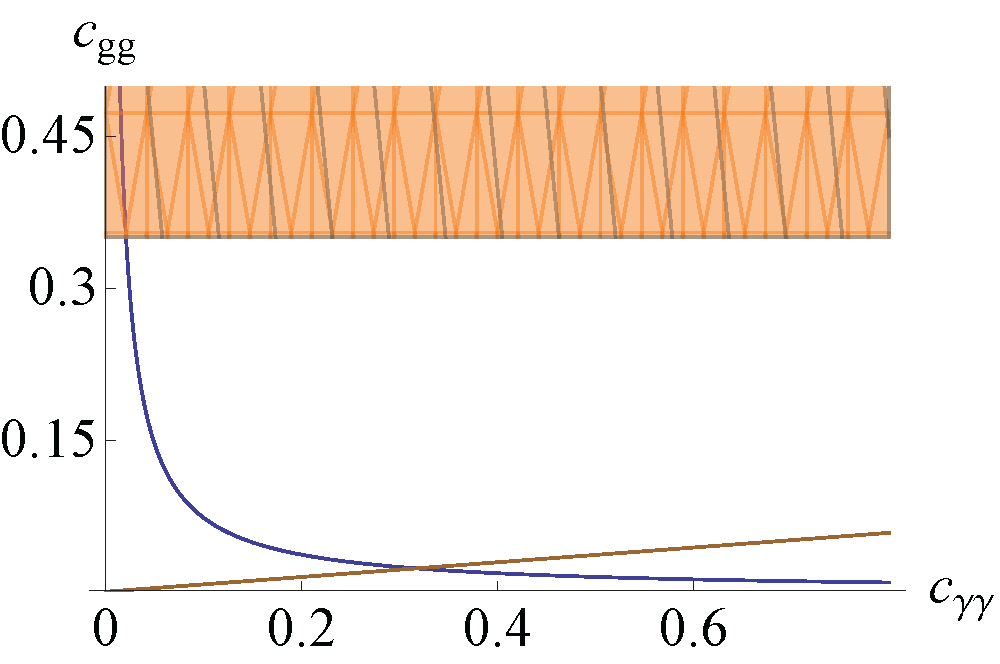}{0.9}
\end{minipage}
\caption{Allowed parameter space on the $c_{gg}$ vs $c_{\xg\xg}$ plane
  for $M_s = 7~{\rm TeV}$ (left) and $M_s = 12~{\rm TeV}$. The blue
  curves indicate the best fit of diphoton cross section ($\xs_{\rm
    LHC13} = 5~{\rm fb}$) for $M_s = 7~{\rm TeV}$ (left) and $M_s =
  12~{\rm TeV}$ (right). The (red and orange) shaded regions are
  excluded at 95\% CL by null results on dijet searches. The slanted
  (brown) curve determines the transition between $c_{gg}$ and
  $c_{\gamma \gamma}$ dominance at production. Gluon fusion dominates
  above this curve. Results for three possible total widths are
  exhibited in the vertical columns, $\xG_{\zt{total}} = 45, 10,
  0.1$~GeV, downwards.}
\label{fig5}
\end{figure}

The assumed coupling of $\varphi$ to the hypercharge field strength yields
additional decay channels in the visible sector, namely $\varphi \to \gamma
Z$ and $\varphi \to ZZ$, with
\begin{equation}
\frac{\Gamma_{\gamma Z}}{\Gamma_{\gamma \gamma}} = 2 \tan^2 \theta_W
\approx 0.6 \quad {\rm and} \quad \frac{\Gamma_{Z Z}}{\Gamma_{\gamma
    \gamma}} =  \tan^4 \theta_W \approx 0.08 \,.
\end{equation}
This prediction is  in agreement with the recent upper limit
reported  in by the ATLAS Collaboration from searches in
the $\gamma Z$ channel~\cite{ATLASzgamma}.
 
We now turn to discuss the production via gluon fusion. We parametrize
the coupling of $\varphi$ to the gluon by the following vertex
\begin{equation}
\frac{c_{gg}}{M_s} \, \varphi \, G^2 .
\end{equation}
The partial decay width of $\xvp$ to dijets is
\begin{equation}
\Gamma_{g g} = 8 \frac{c_{g g}^2}{4\pi}
\frac{M_\xvp^3}{M_s^2}\, . \label{eq:gluonwidth}
\end{equation}
In Fig.~\ref{fig3} we show a scan of the parameter space $(c_{gg},
M_s)$ for constant values of 
 $\Gamma_{g g}$ as obtained from (\ref{eq:gluonwidth}).

In the narrow width approximation the cross section for
diphoton production via gluon fusion is given by~\cite{Csaki:2016raa}
\be \sigma_{\rm
  LHC13} (gg \to \varphi \to \gamma \gamma) = 5.8\times
10^{3}~\text{pb}~c_{gg}^2\lb\frac{M_s}{ \text{TeV}}\rb^{-2} {\cal
  B}(\xvp \rightarrow \gamma\gamma)\, 
\label{LHC13gggaga}
\ee and \be \sigma_{\rm LHC8}
(gg \to \varphi \to \gamma \gamma) = 1.2 \times
10^{3}~\text{pb}~c_{gg}^2\lb\frac {M_s} {\text{TeV}} \rb^{-2} {\cal
  B}(\xvp \rightarrow \gamma\gamma)\, .
\label{LHC8gggaga}
\ee Substituting (\ref{mattS}) and (\ref{shl-2}) into
(\ref{LHC13gggaga}) we arrive at the targeting constraint equation
connecting $c_{gg}$, $c_{\gamma \gamma}$, and $M_s$, for given
$\Gamma_{\rm total}$. In Fig.~\ref{fig4} we show contours of constant
string scale $M_s$ for the best fit of the diphoton cross section
($\xs_{\rm LHC13} = 5~{\rm fb}$) produced via gluon fusion,  with 
different values of the total decay width, $\Gamma_{\rm total} = 45,
10, 0.1~{\rm GeV}$.  By comparing the different panels of the figure
we can see that the phase space critically shrinks with decreasing
$\Gamma_{\rm total}$. For $\varphi$ production via gluon fusion, there
is an additional constraint due to the null result on dijet searches
above SM expectations. It is this that we now turn to study.

As of today the upper limit on dijet production at $M_{jj} = 750~{\rm
  GeV}$ is dominated by LHC8 data, $\sigma_{\rm LHC8} (pp \to jj) <
2.5~{\rm pb}$ at 95\%CL~\cite{Aad:2014aqa}.  The cross section for
dijet production is \be \sigma_{\rm LHC8} (gg \to \varphi \to gg) =
7.6\times 10^{3}~\text{pb}~c_{gg}^4\lb\frac {M_s} {\text{TeV}}
\rb^{-4} \lb\frac {\Gamma_{\zt{total}}} {45~\text{GeV}} \rb\,
. \label{LHC8dijet}\ee Imposing the dijet constraint, $\sigma_{\rm
  LHC8} (pp \to jj) < 2.5~{\rm pb}$, on (\ref{LHC8dijet}) we obtain
the excluded region of the ($c_{gg}, c_{\gamma \gamma}$) plane. The allowed region of the ($c_{gg}, c_{\gamma \gamma}$)
parameter space, which explains
the observed diphoton excess at LHC13 and is consistent with LHC8
data, is shown in Fig.~\ref{fig5} for illustrative
values of the string scale, $M_s = 7~{\rm TeV}$ and 12~{\rm TeV}. From (\ref{LHC13gggaga})
and (\ref{LHC8gggaga}) the LHC13/LHC8 luminosity ratio is found to be
\begin{equation}
  \frac{{\cal L}_{g g} (\sqrt{s} = 13~{\rm TeV})}{{\cal L}_{g g} (\sqrt{s} =
    8~{\rm TeV})} = 4.7.
\end{equation}
Therefore, the production of $\varphi$ by gluon fusion is consistent
with the lack of a diphoton excess in LHC8 data. As we have seen,
imposing the LHC8 dijet limit~\cite{Aad:2014aqa} further constrains
the $(c_{gg},c_{\gamma \gamma})$ parameter space. However, it remains
the case that for $7 \alt M_s/{\rm TeV} \alt 30$ there is always some
allowed region of the ($c_{gg}, c_{\gamma \gamma})$ plane.

Summarizing, we embrace this joyous moment that appears to be the
emergence of non-standard model physics in collider data, by
investigating low-mass-scale string compactifications endowed with
D-brane configurations that realize the SM by open strings. We have
shown that generic D-brane constructs can explain the peak in the
diphoton invariant mass spectrum at 750 GeV recently reported by the
LHC experiments. Under reasonable assumptions, we have demonstrated
that the excess could originate from a closed string (possibly
axionic) excitation $\varphi$ that has a coupling with gauge kinetic
terms. We estimated the $\varphi$ production rate from photon and
gluon fusion. For string scales above todays lower limit $M_s \approx
7~{\rm TeV}$, we can accommodate the diphoton rate observed at Run II
while maintaining consistency with Run I data.

\acknowledgments{L.A.A.  is supported by U.S. National Science
  Foundation (NSF) CAREER Award PHY1053663 and by the National
  Aeronautics and Space Administration (NASA) Grant No. NNX13AH52G; he
  thanks the Center for Cosmology and Particle Physics at New York
  University for its hospitality.  H.G. and
  T.R.T. are supported by NSF Grant No. PHY-1314774.  X.H.  is
  supported by the MOST Grant 103-2811-M-003-024. D.L. is partially
  supported by the ERC Advanced Grant Strings and Gravity
  (Grant.No. 32004) and by the DFG cluster of excellence ``Origin and
  Structure of the Universe.''  Any opinions, findings, and
  conclusions or recommendations expressed in this material are those
  of the authors and do not necessarily reflect the views of the
  National Science Foundation.}


\begin{thebibliography}{99}


\bibitem{ATLAS} 
  ATLAS Collaboration,
  {\color{rossoCP3} Search for resonances decaying to photon pairs in 3.2 fb$^{-1}$ of $pp$ collisions at $\sqrt{s}$ = 13 TeV with the ATLAS detector},
  ATLAS-CONF-2015-081.


\bibitem{CMS:2015dxe} 
  CMS Collaboration,
  {\color{rossoCP3} Search for new physics in high mass diphoton events in proton-proton
  collisions at 13~TeV},
  CMS-PAS-EXO-15-004.


\bibitem{Franceschini:2015kwy} 
  R.~Franceschini {\it et al.},
  arXiv:1512.04933 [hep-ph].


\bibitem{Aad:2015mna} 
  G.~Aad {\it et al.} [ATLAS Collaboration],
  {\color{rossoCP3} Search for high-mass diphoton resonances in $pp$ collisions at $\sqrt{s}=8$ TeV with the ATLAS detector},
  Phys.\ Rev.\ D {\bf 92}, 032004 (2015)
  doi:10.1103/PhysRevD.92.032004
  [arXiv:1504.05511 [hep-ex]].

\bibitem{Khachatryan:2015qba} 
  V.~Khachatryan {\it et al.} [CMS Collaboration],
  {\color{rossoCP3} Search for diphoton resonances in the mass range from 150 to 850~GeV in $pp$ collisions at $\sqrt{s} =$ 8~TeV},
  Phys.\ Lett.\ B {\bf 750}, 494 (2015)
  doi:10.1016/j.physletb.2015.09.062
  [arXiv:1506.02301 [hep-ex]].

\bibitem{Anchordoqui:2015jxc} 
  L.~A.~Anchordoqui, I.~Antoniadis, H.~Goldberg, X.~Huang, D.~L\"ust and T.~R.~Taylor,
  {\color{rossoCP3} 750 GeV diphotons from closed string states},
  Phys.\ Lett.\ B {\bf 755}, 312 (2016)
  doi:10.1016/j.physletb.2016.02.024
  [arXiv:1512.08502 [hep-ph]].



\bibitem{Antoniadis:1998ig} 
  I.~Antoniadis, N.~Arkani-Hamed, S.~Dimopoulos and G.~R.~Dvali,
  {\color{rossoCP3} New dimensions at a millimeter to a Fermi and superstrings at a TeV},
  Phys.\ Lett.\ B {\bf 436}, 257 (1998)
  doi:10.1016/S0370-2693(98)00860-0
  [hep-ph/9804398].




\bibitem{Blumenhagen:2005mu}
  R.~Blumenhagen, M.~Cvetic, P.~Langacker and G.~Shiu,
  {\color{rossoCP3} Toward realistic intersecting D-brane models},
  Ann.\ Rev.\ Nucl.\ Part.\ Sci.\  {\bf 55}, 71 (2005)
  doi:10.1146/annurev.nucl.55.090704.151541
  [hep-th/0502005].



\bibitem{Blumenhagen:2006ci} 
  R.~Blumenhagen, B.~Kors, D.~L\"ust and S.~Stieberger,
  {\color{rossoCP3} Four-dimensional string compactifications with D-branes, orientifolds and fluxes},
  Phys.\ Rept.\  {\bf 445}, 1 (2007)
  doi:10.1016/j.physrep.2007.04.003
  [hep-th/0610327].







\bibitem{Heckman:2015kqk}
  J.~J.~Heckman,
  {\color{rossoCP3} 750~GeV diphotons from a D3-brane},
  arXiv:1512.06773 [hep-ph].



\bibitem{Cvetic:2015vit} 
  M.~Cvetic, J.~Halverson and P.~Langacker,
  {\color{rossoCP3} String consistency, heavy exotics, and the $750$~GeV diphoton excess at the LHC},
  arXiv:1512.07622 [hep-ph].


\bibitem{Ibanez:2015uok} 
  L.~E.~Ibanez and V.~Martin-Lozano,
   {\color{rossoCP3}  A megaxion at 750~GeV as a first hint of low scale string theory},
  arXiv:1512.08777 [hep-ph].


\bibitem{Palti:2016kew} 
  E.~Palti,
   {\color{rossoCP3} Vector-like exotics in F-theory and 750~GeV diphotons},
  arXiv:1601.00285 [hep-ph].


\bibitem{Karozas:2016hcp}
  A.~Karozas, S.~F.~King, G.~K.~Leontaris and A.~K.~Meadowcroft,
   {\color{rossoCP3} 750 GeV diphoton excess from $E_6$ in F-theory GUTs},
  arXiv:1601.00640 [hep-ph].



\bibitem{Faraggi:2016xnm} 
  A.~E.~Faraggi and J.~Rizos,
   {\color{rossoCP3} The 750 GeV diphoton LHC excess and extra $Z'$s in heterotic-string derived models},
  arXiv:1601.03604 [hep-ph].





\bibitem{Anastasopoulos:2016cmg} 
  P.~Anastasopoulos and M.~Bianchi,
   {\color{rossoCP3} Revisiting light stringy states in view of the 750~GeV diphoton excess},
  arXiv:1601.07584 [hep-th].


\bibitem{Cvetic:2016omj} 
  M.~Cvetic, J.~Halverson and P.~Langacker,
   {\color{rossoCP3} String consistency, heavy exotics, and the 750~GeV diphoton excess at the LHC: Addendum},
  arXiv:1602.06257 [hep-ph].


\bibitem{Li:2016tqf} 
  T.~Li, J.~A.~Maxin, V.~E.~Mayes and D.~V.~Nanopoulos,
   {\color{rossoCP3} The 750~GeV diphoton excesses in a realistic D-brane model},
  arXiv:1602.09099 [hep-ph].




\bibitem{Leonatris} 
  G.~K.~Leontaris and Q.~Shafi,
   {\color{rossoCP3}  Diphoton resonance in F-theory inspired flipped $SO(10)$},
  arXiv:1603.06962 [hep-ph].




\bibitem{Abel:2016pyc} 
S.~Abel and V.~V.~Khoze, {\color{rossoCP3} Photo-production of a
    750 GeV di-photon resonance mediated by Kaluza-Klein leptons in
    the loop}, arXiv:1601.07167 [hep-ph].




\bibitem{Higaki:2015jag} 
  T.~Higaki, K.~S.~Jeong, N.~Kitajima and F.~Takahashi,
   {\color{rossoCP3} The QCD axion from aligned axions and diphoton excess},
  Phys.\ Lett.\ B {\bf 755}, 13 (2016)
  doi:10.1016/j.physletb.2016.01.055
  [arXiv:1512.05295 [hep-ph]].




\bibitem{Megias:2015ory} 
  E.~Megias, O.~Pujolas and M.~Quiros,
   {\color{rossoCP3} On dilatons and the LHC diphoton excess},
  arXiv:1512.06106 [hep-ph].




\bibitem{Ben-Dayan:2016gxw} 
  I.~Ben-Dayan and R.~Brustein,
   {\color{rossoCP3} Hypercharge axion and the diphoton $750$~GeV resonance},
  arXiv:1601.07564 [hep-ph].



\bibitem{Barrie:2016ntq} 
  N.~D.~Barrie, A.~Kobakhidze, M.~Talia and L.~Wu,
   {\color{rossoCP3} 750 GeV composite axion as the LHC diphoton resonance},
  Phys.\ Lett.\ B {\bf 755}, 343 (2016)
  doi:10.1016/j.physletb.2016.02.010
  [arXiv:1602.00475 [hep-ph]].




\bibitem{Aparicio:2016iwr} 
  L.~Aparicio, A.~Azatov, E.~Hardy and A.~Romanino,
   {\color{rossoCP3} Diphotons from diaxions},
  arXiv:1602.00949 [hep-ph].








\bibitem{Delmastro} M. Delmastro [on behalf of the ATLAS Collaboration],
  {\color{rossoCP3} Diphoton searches in ATLAS},
51st Rencontres de Moriond (Electroweak session) 17 May 2016, La Thuile (Italy).

\bibitem{Musella} P. Musella [on behalf of the CMS Collaboration],
  {\color{rossoCP3} Search for high mass diphoton resonances at CMS},
51st Rencontres de Moriond (Electroweak session) 17 May 2016, La Thuile (Italy).

\bibitem{CMS:2016owr} 
  CMS Collaboration,
   {\color{rossoCP3}  Search for new physics in high mass diphoton events in $3.3~\mathrm{fb}^{-1}$ of proton-proton collisions at $\sqrt{s}=13~\mathrm{TeV}$ and combined interpretation of searches at $8~\mathrm{TeV}$ and $13~\mathrm{TeV}$},
  CMS-PAS-EXO-16-018.




\bibitem{Csaki:2016raa} 
  C.~Csaki, J.~Hubisz, S.~Lombardo and J.~Terning,
    {\color{rossoCP3} Gluon vs. Photon Production of a 750 GeV Diphoton Resonance},
  arXiv:1601.00638 [hep-ph].



\bibitem{Harland-Lang:2016qjy} 
  L.~A.~Harland-Lang, V.~A.~Khoze and M.~G.~Ryskin,
    {\color{rossoCP3} The production of a diphoton resonance via photon-photon fusion},
  arXiv:1601.07187 [hep-ph].

\bibitem{Martin:2016byf} 
  A.~D.~Martin and M.~G.~Ryskin,
   {\color{rossoCP3} Advantages of exclusive $\gamma \gamma$ production to probe high mass systems},
  J.\ Phys.\ G {\bf 43}, no. 4, 04LT02 (2016)
  doi:10.1088/0954-3899/43/4/04LT02
  [arXiv:1601.07774 [hep-ph]].

\bibitem{Kats:2016kuz} 
  Y.~Kats and M.~Strassler,
 {\color{rossoCP3}  Resonances from QCD bound states and the 750 GeV diphoton excess},
  arXiv:1602.08819 [hep-ph].




\bibitem{Khachatryan:2015dcf} 
  V.~Khachatryan {\it et al.} [CMS Collaboration],
  {\color{rossoCP3} Search for narrow resonances decaying to dijets in proton-proton collisions at $\sqrt{s} = 13$~TeV},
  arXiv:1512.01224 [hep-ex].


\bibitem{Antoniadis:2002cs} 
  I.~Antoniadis, E.~Kiritsis and J.~Rizos,
  {\color{rossoCP3} Anomalous $U(1)$'s in type 1 superstring vacua},
  Nucl.\ Phys.\ B {\bf 637}, 92 (2002)
  doi:10.1016/S0550-3213(02)00458-3
  [hep-th/0204153].




\bibitem{ATLAS-let} ATLAS Collaboration, 
{\color{rossoCP3} Letter of Intent for the Phase-I Upgrade of the
  ATLAS Experiment}, Technical Report CERN- LHCC-2011-012. LHCC-I-020,
  CERN, Geneva, Nov. 2011.

\bibitem{Albrow} M Albrow {\it et al.} [CMS and TOTEM Collaborations],
 {\color{rossoCP3} CMS-TOTEM Precision Proton Spectrometer}, Technical Report CERN-LHCC-2014-021. TOTEM-TDR-003. CMS-TDR-13, CERN, Geneva, Sep. 2014.

\bibitem{Dienes:2011ja} 
  K.~R.~Dienes and B.~Thomas,
  {\color{rossoCP3} Dynamical dark matter: I. Theoretical overview},
  Phys.\ Rev.\ D {\bf 85}, 083523 (2012)
  doi:10.1103/PhysRevD.85.083523
  [arXiv:1106.4546 [hep-ph]].


\bibitem{Dienes:2011sa} 
  K.~R.~Dienes and B.~Thomas,
 {\color{rossoCP3} Dynamical dark matter: II. An explicit model},
  Phys.\ Rev.\ D {\bf 85}, 083524 (2012)
  doi:10.1103/PhysRevD.85.083524
  [arXiv:1107.0721 [hep-ph]].

\bibitem{ATLASzgamma} 
  ATLAS Collaboration,
 {\color{rossoCP3}  Search for heavy resonances decaying to a $Z$ boson and a photon in $pp$ collisions at $\sqrt{s}=13$ TeV with the ATLAS detector},
  ATLAS-CONF-2016-010.



\bibitem{Schmidt:2015zda} 
  C.~Schmidt, J.~Pumplin, D.~Stump and C.-P.~Yuan,
  {\color{rossoCP3} CT14QED PDFs from Isolated Photon Production in Deep Inelastic Scattering},
  arXiv:1509.02905 [hep-ph].


\bibitem{Martin:2004dh} 
  A.~D.~Martin, R.~G.~Roberts, W.~J.~Stirling and R.~S.~Thorne,
  {\color{rossoCP3} Parton distributions incorporating QED contributions},
  Eur.\ Phys.\ J.\ C {\bf 39}, 155 (2005)
  doi:10.1140/epjc/s2004-02088-7
  [hep-ph/0411040].

\bibitem{Ball:2013hta} 
  R.~D.~Ball {\it et al.} [NNPDF Collaboration],
  {\color{rossoCP3} Parton distributions with QED corrections},
  Nucl.\ Phys.\ B {\bf 877}, 290 (2013)
  doi:10.1016/j.nuclphysb.2013.10.010
  [arXiv:1308.0598 [hep-ph]].












\bibitem{Aad:2014aqa} 
  G.~Aad {\it et al.} [ATLAS Collaboration],
  {\color{rossoCP3} Search for new phenomena in the dijet mass distribution using $pp$ collision data at $\sqrt{s}=8$~TeV with the ATLAS detector},
  Phys.\ Rev.\ D {\bf 91},  052007 (2015)
  doi:10.1103/PhysRevD.91.052007
  [arXiv:1407.1376 [hep-ex]].






\end{thebibliography}
\end{document}